\documentclass[10pt,journal,draftclsnofoot,onecolumn]{IEEEtran}
\usepackage{paralist}
\usepackage{hyperref}
\usepackage{enumerate}
\usepackage{url}
\usepackage{amsthm}

\usepackage[pdftex]{graphicx}
\usepackage{algpseudocode}
\usepackage{algorithm}
\usepackage{amsmath}
\usepackage{amssymb}
\usepackage{upref}
\usepackage{amsfonts}
\usepackage{graphicx}
\usepackage{epic}
\usepackage{psfrag}
\usepackage{verbatim}
\usepackage{amsfonts}
\usepackage{amssymb}
\usepackage{array}
\usepackage{color}
\usepackage{textcomp}

\newcommand{\cG}{{\cal G}}

\newcommand{\cP}{{\cal P}}

\newcommand{\sP}{\cP}
\newcommand{\sG}{\cG}

\newcommand{\Gr}{\smash{{\sG\kern-1.5pt}_q\kern-0.5pt(n,k)}}
\newcommand{\Grr}{\smash{{\sG\kern-1.5pt}_q\kern-0.5pt(n,r)}}
\newcommand{\Gfourk}{\smash{{\sG\kern-1.5pt}_q\kern-0.5pt(4k,2k)}}
\newcommand{\Gk}{\smash{{\sG\kern-1.5pt}_q\kern-0.5pt(n,k_1)}}
\newcommand{\Gkk}{\smash{{\sG\kern-1.5pt}_q\kern-0.5pt(n,k_2)}}
\newcommand{\Grtwo}{\smash{{\sG\kern-1.5pt}_2\kern-0.5pt(n,k)}}
\newcommand{\Gkone}{\smash{{\sG\kern-1.5pt}_q\kern-0.5pt(n,k_1)}}
\newcommand{\Gktwo}{\smash{{\sG\kern-1.5pt}_q\kern-0.5pt(n,k_2)}}
\newcommand{\Ps}{\smash{{\sP\kern-2.0pt}_q\kern-0.5pt(n)}}

\newtheorem{theorem}{Theorem}

\newtheorem{definition}[theorem]{Definition}

\begin{document}
\title{Local Rank Modulation for Flash Memories}
\author{Michal Horovitz
\thanks{M. Horovitz is with the Department of Computer Science,
Technion --- Israel Institute of Technology, Haifa 32000, Israel.
(email: michalho@cs.technion.ac.il). This work is part of her M.Sc.
thesis performed at the Technion.}
}
\maketitle
\begin{abstract}
Local rank modulation scheme was suggested recently for
representing information in flash memories 
in order to overcome drawbacks of rank modulation.
For $s\leq t\leq n$ with $s|n$, $(s,t,n)$-LRM scheme
is a local rank modulation scheme 
where the $n$ cells are locally viewed through a
sliding window of size $t$ 
resulting in a sequence of small permutations which
requires less comparisons and less distinct values.
The distance between two windows equals to~$s$.
To get the simplest hardware implementation 
the case of sliding window of size two was presented.
Gray codes and constant weight Gray codes were presented in order 
to exploit the full representational power of the scheme.
In this work, a tight upper-bound for cyclic constant weight Gray code 
in $(1,2,n)$-LRM scheme where the weight equals to $2$ is given.
Encoding, decoding and enumeration of 
$(1,3,n)$-LRM scheme is studied.
\end{abstract}


\maketitle
\section{Introduction}
Flash memory is a non-volatile technology that is both electrically
programmable and electrically erasable. It incorporates a set of cells
maintained at a set of levels of charge to encode information.
While raising the charge level of a cell is an easy operation,
reducing the charge level requires the erasure of the whole block to which the cell belongs.
For this reason charge is injected into the cell over several iterations.
Such programming is slow and can cause errors since cells may be injected with extra unwanted charge.
Other common errors in flash memory cells are due to charge leakage and reading disturbance that
may cause charge to move from one cell to its adjacent cells.
In order to overcome these problems, the novel 
framework of \emph{rank modulation} was introduced in~\cite{rankModulation1_2009}.
In this setup the information is carried by the relative ranking of the
cells' charge levels and not by the absolute values of the charge levels.
This allows for more efficient programming of cells, and coding by the ranking of the cells' charge levels
is more robust to charge leakage than coding by their actual values.
The \emph{push-to-the-top} operation is a basic minimal cost
operation in the rank modulation scheme by which a single cell has its charge
level increased so as to be the highest of the set.

A drawback of the rank modulation is the need for a large number of
comparisons when reading the induced permutation. Furthermore, distinct
$n$ charge levels are required for a group of~$n$ cells. 
The \emph{local rank modulation} scheme was suggested in order to overcome these problems.
In this scheme, the $n$ cells are locally viewed through a
sliding window, resulting in a sequence of small permutations which
requires less comparisons and less distinct values. 
For $s\leq t\leq n$ with $s|n$, $(s,t,n)$-LRM scheme, defined in \cite{localRankModulation1_2011}, 
is a local rank modulation scheme over~$n$ physical cells, 
where~$t$ is the size of each sliding window, and~$s$ is the distance between two windows.
In this scheme, the push-to-the-top operation 
merely raises the charge level of the selected cell 
above those cells which are comparable with it.
We say a sequence~$f$ of $n/s$ permutations from $S_t$ is
$(s,t,n)$-LRM scheme \emph{realizable} if it can be demodulated 
to a sequence of charges in~$n$ cells under $(s,t,n)$-LRM scheme. 
Except for the degenerate case where $s=t=n$, not every sequence is realizable.

In this paper we discuss two topics.
In Section \ref{sec:12nScheme} 
we introduce a tight upper-bound for $(1,2,n;2)$-LRMGC 
(a constant-weight-Gray-code in $(1,2,n)$-LRM scheme where the weight is $2$),
and in Section \ref{sec:13nScheme}, $(1,3,n)$-LRM scheme is studied.

\section{The $(1,2,n)$-LRM scheme}
\label{sec:12nScheme}
$(1,2,n)$-LRM scheme is a local rank modulation scheme over $n$ physical cells, where the size
of each sliding window is $2$, and each cell starts a new window.
Thus, only two permutations exist:
$[1,2]$ associated with the logical value $1$, and $[2,1]$ associated with $0$. 
Therefore, in $(1,2,n)$-LRM scheme we store about one bit per cell, 
which requires just one comparison per cell for reading, 
and perform comparisons with two cells for a push-to-the-top operation.
It is easily verified that the only two binary sequences not mapped to $(1,2,n)$-LRM scheme
are the all-ones and all-zeros sequences.
Hence, the set of the realizable words in $(1,2,n)$-LRM scheme
is $S(n)=\{0,1\}^n\setminus\{0^n,1^n\}$.
The push-to-the-top operation raises the charge level of
the selected cell above its adjacent cells, 
therefore it is made by selecting a window of size $2$ 
in the original codeword and overwriting it with $01$.

In \cite{rankModulation1_2009} Gray codes were
presented in order to exploit the full representational power
of the rank modulation scheme and data rewriting schemes.

\begin{definition}
A Gray code, $G$, for $(1,2,n)$-LRM scheme (denoted by
$(1,2,n)$-LRMGC) is a sequence of $N$ distinct length~$n$ binary
codewords from $S(n)$. 
$G=g_0, g_1, \ldots, g_{N-1}$, where for each
$0\leq i\leq N-2$, $g_{i+1}$ is a result of a push-to-the-top operation on $g_i$ 
If $g_0$ is also a result of a push-to-the-top operation on $g_{N-1}$ then we say that $G$ is
cyclic.
\end{definition}

The weight of $g$ where $g\in \{0,1\}^n$, 
denoted by $w(g)$, is the number of $1$'s in $g$.
Let $S(n,w)$ be the set of all codewords in $S(n)$ with weight $w$.

\begin{definition}
Let ${G=g_0,g_1\ldots,g_{N-1}}$ be a Gray code 
for $(1,2,n)$-LRM scheme.
$G$ is a constant-weight Gray code for $(1,2,n)$-LRM scheme
(denoted by $(1,2,n;w)$-LRMGC) if for each
$0\leq i\leq N-1$, $g_i\in S(n,w)$.
\end{definition}

The motivation for constant-weight Gray codes was described in \cite{localRankModulation1_2011}.
The transitions between adjacent words in the constant-weight
variant of $(1,2,n)$-LRM scheme replace a window of size $2$ in $g_i$ which contains $10$ 
with $01$ in $g_{i+1}$, i.e., 'pushing' of a logical '$1$' a single place to the right. 

\subsection{Upper bound for the size of a cyclic $(1,2,n;2)$-LRMGC}
Let $C$ be a cyclic $(1,2,n;2)$-LRMGC of size $N$.
The obvious question to be asked is, what is the size of the largest code $C$?
It was proved in \cite{localRankModulation1_2011} (Theorem 8) 
that ${N\leq \binom{n}{2}-\frac{1}{8}(n-3)(n-5)}$.
This bound was obtained by translating the question into a graph problem.
The related graph $\mathcal{G}_n$ is defined in \cite{localRankModulation1_2011}. 
The set of vertices in $\mathcal{G}_n$ is $S(n,2)$,
and there exists an edge $v\to v'$ in $\mathcal{G}_n$ if and only if~$v'$ can follow~$v$ in a $(1,2,n;2)$-LRMGC.
By a careful analysis of~$\mathcal{G}_n$, we obtain the following result.
\begin{theorem}
If $C$ is a cyclic $(1,2,n;2)$-LRMGC of size~$N$ then $N\leq 2n$.
\end{theorem}
The complete (long) proof will be given in the full version of this paper.
Obviously there exists a cyclic $(1,2,n;w)$-LRMGC of size $2n$ (see \cite{localRankModulation1_2011}).
Thus we have that $2n$ is a tight upper-bound.

\section{The $(1,3,n)$-LRM scheme}
\label{sec:13nScheme}
$(1,3,n)$-LRM scheme is a local rank modulation scheme over~$n$ physical cells, 
where the size of each sliding window is~$3$, and each cell starts a new window.
Since the size of a sliding window is~$3$, 
demodulated sequences of permutations in this scheme contain $3!$ permutations. 
Therefore we need an alphabet of size $6$ 
to present the demodulated sequences of permutations. 
The alphabet $S=\{0,1,\ldots, 5\}$ represents $6$ permutations as follows.
\begin{center}
\begin{tabular}{lll}
$0$ & $\triangleq$ & $[1,2,3]$,\\
$1$ & $\triangleq$ & $[1,3,2]$,\\
$2$ & $\triangleq$ & $[2,1,3]$,\\
$3$ & $\triangleq$ & $[3,1,2]$,\\
$4$ & $\triangleq$ & $[2,3,1]$,\\
$5$ & $\triangleq$ & $[3,2,1]$.
\end{tabular}
\end{center}

We denote the words over this alphabet as \emph{base-words},
and define a mapping of the base-words
to \emph{codewords} over an alphabet of size $3$.

Let $S^e=\{0,2,4\}$ and $S^o=\{1,3,5\}$ 
be a partition of $S$ into \emph{even} and \emph{odd} symbols, respectively.
Let ${\alpha=(\alpha_0,\alpha_1,\ldots, \alpha_{n-1})}$ be a base-word. 
Note that the last two cells which determine $\alpha_i$ are the first two cells which represent $\alpha_{i+1}$, 
where $0\leq i\leq n-1$ and $i+1$ is taken modulo $n$.
Therefore, given $\alpha_i$, there are only three options for $\alpha_{i+1}$.
Let $\tilde{S^e}$ and $\tilde{S^o}$ be the sets of symbols that
can follow the symbols in $S^e$ and $S^o$, respectively.
It can be easily verified that $\tilde{S^e}=\{0,1,3\}$ and $\tilde{S^o}=\{2,4,5\}$.

The base-word $\alpha$ is mapped to a codeword
$c=(c_0,c_1,\ldots,c_{n-1})$.
The relation between $\alpha_i$, $\alpha_{i+1}$, 
and $c_i$ where $0\leq i\leq n-1$ and $i+1$ is taken modulo~$n$, 
is presented in the following table.
\begin{center}
\begin{tabular}{|l||l|l|l|}
\hline
$\alpha_i\in S^e$ & $\alpha_{i+1}=0$ & $\alpha_{i+1}=1$ & $\alpha_{i+1}=3$ \\
\hline
$\alpha_i\in S^o$ & $\alpha_{i+1}=2$ & $\alpha_{i+1}=4$ & $\alpha_{i+1}=5$ \\
\hline
\hline
 & $c_i=0$ & $c_i=1$ & $c_i=2$ \\
\hline
\end{tabular}
\end{center}

A length~$n$ codeword, $c$, over the alphabet $\{0,1,2\}$ is \emph{legal}, 
if there exists a realizable base-word $\alpha$, such that $\alpha$ is encoded to $c$.
Note that not all the base-words are realizable.
The last cell is compared with the first two cells, 
and the cell before is compared  with the first cell.
Thus, there exists a base-word, $\alpha=(\alpha_0,\alpha_1,\ldots, \alpha_{n-1})$, 
such that $\alpha$ satisfies the dependence between $\alpha_{i+1}$ and $\alpha_i$ 
(for each $0\leq i\leq n-1$ where $i+1$ is taken modulo $n$),
but $\alpha$ is still not realizable.
For example, the following base-words are not realizable:
\begin{itemize}
\item $0^n$ - the charge levels are always decreased.
\item $5^n$ - the charge levels are always increased.
\item $(14)^{(n/2)}$ (where $n$ is even) - the charge levels are decreased in the odd cells
and increased in the even cells, and the levels of the odd cells are always higher than the levels in the even cells.
\end{itemize}

The only two base-words mapped to the codeword $1^n$ are $(14)^{(n/2)}$ and $(41)^{(n/2)}$.
These base-words are not realizable.
Therefore the all-ones codeword is not legal.

Thus, given a legal codeword $c=(c_0,c_1,\ldots,c_{n-1})$, 
there exists $0\leq i \leq n-1$, such that $c_i\in \{0,2\}$.
Without loss of generality, we can assume that $c_0\ne 1$ 
(since the base-words and the codewords are cyclic).

If $c_0=0$ then we have $\alpha_1\in \{0,2\}$, i.e., $\alpha_1$ is even.
Thus, $\alpha_2$ is determined by an entry in the \emph{first} row in the above table, 
where the column is chosen according to $c_2$.
If $c_0=2$ then we have $\alpha_1\in \{3,5\}$, i.e., $\alpha_1$ is odd.
Thus, $\alpha_2$ is determined by an entry in the \emph{second} row in the above table, 
where the column is chosen according to $c_2$.

Now, it is easy to determine $\alpha_3,\alpha_4,\ldots , \alpha_{n-1}$
and also $\alpha_0$ and $\alpha_1$.
Note that if $\alpha_1$ is not equal
to an optional initial value 
(from the set $\{0,2\}$ if $c_0=0$ and from $\{3,5\}$ if $c_0=2$)
then we can conclude that $c$ is not legal.

This method provides us an one-to-one mapping 
between the realizable base-words and the legal codewords.
But, also some non-realizable base-words (due to the charge levels) are mapped to codewords.
Therefore these codeword are illegal.
Thus, decoding a given codeword to a base-word 
doesn't guarantee that the codeword is legal.
For example, the base-word ${\alpha=5^n}$ is mapped to the codeword ${c=2^n}$ 
and ${\alpha=0^n}$ is mapped to ${c=0^n}$.

The number of legal codewords is exactly 
the number of the realizable base-words.
This number can be obtained 
by constructing $33$ recursive equations 
which describe the relations between the charge levels of the last two and the first two cells.
A careful analysis of these $33$ equations yields the following result 
which provides the motivation for using this scheme.
\begin{theorem}
If $M$ is the number of legal words in $(1,3,n)$-LRM scheme
then $\lim\limits_{n\to \infty}\frac{M}{3^n}=1$.
\end{theorem}

Some of the results on $(1,3,n)$-LRM scheme can be generalized to $(1,t,n)$-LRM scheme for each $t>3$.
It is currently under researched if for $M_t$, the number of legal words in $(1,t,n)$-LRM scheme,
we have $\lim\limits_{n\to \infty}\frac{M_t}{t^n}=1$ for each $t>3$.


\begin{thebibliography}{10}
\bibitem{rankModulation1_2009}
A.~Jiang, R.~Mateescu, M.~Schwartz, and J.~Bruck, ``Rank modulation for flash
  memories,'' \emph{IEEE Transactions on Information Theory}, vol.~55, no.~6,
  pp. 2659--2673, 2009.

\bibitem{localRankModulation1_2011}
E.~E. Gad, M.~Langberg, M.~Schwartz, and J.~Bruck, ``Constant-weight gray codes
  for local rank modulation,'' \emph{IEEE Transactions on Information Theory},
  vol.~57, no.~11, pp. 7431--7442, 2011.

\end{thebibliography}
\end{document}